\documentclass[aps, pra, superscriptaddress, showkeys, twocolumn]{revtex4-1}
\usepackage{graphicx, color}
\usepackage{mathtools, amsthm, amsfonts, amsmath}
\usepackage{natbib}
\usepackage{hyperref}
\usepackage[english]{babel}

\begin{document}
\title{Analysis of the current-driven domain wall motion in a ratchet ferromagnetic strip}

\author{L. S\'{a}nchez-Tejerina}
\email[\textit{Email address:}]{luis.st@ee.uva.es}
\affiliation{Dpto. Electricidad y Electr\'{o}nica. Facultad de Ciencias, Universidad de Valladolid, 47011 Valladolid, Spain}

\author{\'{O}. Alejos}
\affiliation{Dpto. Electricidad y Electr\'{o}nica. Facultad de Ciencias, Universidad de Valladolid, 47011 Valladolid, Spain}

\author{E. Mart\'{\i}nez} 
\affiliation{Dpto. Física Aplicada, Facultad de Ciencias, Universidad de Salamanca, 37011 Salamanca, Spain}

\author{V. Raposo}
\affiliation{Dpto. Física Aplicada, Facultad de Ciencias, Universidad de Salamanca, 37011 Salamanca, Spain}

\begin{abstract}
The current-driven domain wall motion in a ratchet memory due to spin-orbit torques is studied from both full micromagnetic simulations and the one dimensional model. Within the framework of this model, the integration of the anisotropy energy contribution leads to a new term in the well known q-$\Phi$ equations, being this contribution responsible for driving the domain wall to an equilibrium position. The comparison between the results drawn by the one dimensional model and full micromagnetic simulations proves the utility of such a model in order to predict the current-driven domain wall motion in the ratchet memory. Additionally, since current pulses are applied, the paper shows how the proper working of such a device requires the adequate balance of excitation and relaxation times, being the latter longer than the former. Finally, the current-driven regime of a ratchet memory is compared to the field-driven regime described elsewhere, then highlighting the advantages of this current-driven regime.\\
\end{abstract}

\keywords{Domain Wall motion, Magnetic memory, Ratchet memory, Magnetocristalline Anisotropy.}

\maketitle

\section*{Introduction}

Domain Walls (DW) are defined as the boundaries between regions of different magnetization (domains) in a magnetic ordered system. DWs in ultrathin films, where high perpendicular anisotropy appears due to this ultra low thickness,\citep{MagAnis} have been recurrently proposed as promising devices for data storage based on DW motion caused by applied fields \citep{DWMobCow, Thia12} or currents. \citep{VanColAp, HDWvelMoore} In particular, the motion of these DWs under the solely application of a magnetic field $H_z$ parallel to the magnetization of the domains is a well-known issue. \citep{WalkerBrd} Under such an excitation, it is known that the velocity of the DW increases linearly until a well defined field, known as Walker field $H_{W}$, is reached. Above this field the DW velocity falls down, reaches a minimum, and then starts increasing, eventually overcoming the maximum speed achieved at $H_{W}$. The most characteristic difference between both regimes, below ($H_z<H_{W}$) and above ($H_z>H_{W}$) the Walker field, is that the former regime is characterized by a constant orientation of the inner magnetization of the DW, while  the latter imposes that this magnetization precesses around the applied magnetic field. This behavior, together with the fact that applied fields drive DWs of different types in opposite directions, constitutes a serious handicap for the implementation of DW-based memory elements.

However, recent studies focused on multilayer systems with inversion asymmetry, where a ferromagnetic layer (FM) is sandwiched between a heavy metal (HM) and an oxide (Ox), \citep{RRashbaSHE, ChiralDWD, CDDWDfullmuMag1D} have open new promising ways of efficient DW-driving. Particularly, the presence of the interfacial Dzyaloshinskii-Moriya interaction (DMI) at the HM/FM interface makes DWs adopt an homochiral N{\'e}el configuration, which is important for such an efficient driving, since DWs can be pushed by the torque exerted by an injected current through the HM, due to the spin Hall effect (SHE). The conclusions drawn by these studies have shown that there is not an analogous {\em Walker current}, and the DW velocity continuously increases up to a value which is reached as the orientation of the magnetization within the DW approaches the direction perpendicular to the current, due to the exerted torque.\\

Once the suitable procedure for DW-driving is provided, a pinning system is required to exactly control the DW position in the FM strip. \citep{FrankenPinning, MDWRM} This problem has been solved in most cases by introducing notches along the strip. However, this tactic was designed for systems with in-plane magnetization, while in PMA systems this effect is weaker, and induce deformations of the DW. For these reasons, some alternatives have been proposed, such as the application of a voltage in an epitaxial magnetic tunnel junction (MTJ), \citep{VoltIndMagAnis} or a strain-mediated coupling in piezoelectric/magnetostrictive bilayer structures. \citep{StrContrAnis} The control of the pinning (and DW nucleation) by means of a tailored PMA has also been proposed. \citep{FrankenPinning} The sample is irradiated with heavy ions so as to create an anisotropy landscape characterized by a sawtooth profile. This idea led to a new proposal for a magnetic memory, known as {\em ratchet memory}, which was studied under the field-driven regime, in particular, the effect of an alternate applied magnetic field with fixed orientation. \citep{RatchMem} The sawtooth was then meant to both fix DW positions (and so define the bit size), and establish one direction of bit shifting by avoiding DWs backward movement due to the applied field.\\

This work goes deeper into this idea, but proposes a current-driven mechanism to ensure the proper bit shifting along the FM strip, which constitutes a much more interesting alternative from the technological point of view to the field-driven basis, since current both promotes the dynamics of all DWs in the same direction and does not lead to precessional DW dynamics, as magnetic field does. Both facts contribute to reduce bit sizes (and so increase bit densities), and speed up bit shifting. The work has been carried out with the help of micromagnetic ($\mu$Mag) simulations as to mimic as much as possible realistic conditions and to explore the feasibility of the proposed mechanisms. Additionally, a one dimensional model (1DM), updated so as to include an effective field accounting for the anisotropy landscape, has been used in order to clarify some aspects of the DW dynamics and can be of help in the further development of the proposed system. According to this, the work is structured as follows. Section~\ref{GeoMod} presents the system under study and describes the 1DM. In Section~\ref{Results} full $\mu$Mag simulations at $T=300$ are presented and compared to those provided by the 1DM. This section focuses on the tuning of the performances of the system by means of the adequate current amplitude, pulse time and relaxation time, being divided in three subsections. While in section~\ref{50C} $50\%$ duty cycle pulses have been used, section~\ref{33C} shows how larger relaxation times improve the system response, by reducing the duty cycle to $33\%$. A short discussion is made in Section~\ref{varDMI} about the effect of not considering a constant DMI parameter, since ion irradiation may also alter the characteristics of this interaction. Additionally, the results shown for the duty cycle of $33\%$ have been used to be compared with those obtained for the field-driven ratchet memory. \citep{RatchMem} The main conclusions of the study are drawn in Section~\ref{concl}.\\

\section{Geometry and models}\label{GeoMod}
\subsection{$\mu$Mag Simulations}

\begin{figure*}[ht]
\includegraphics[scale=0.33]{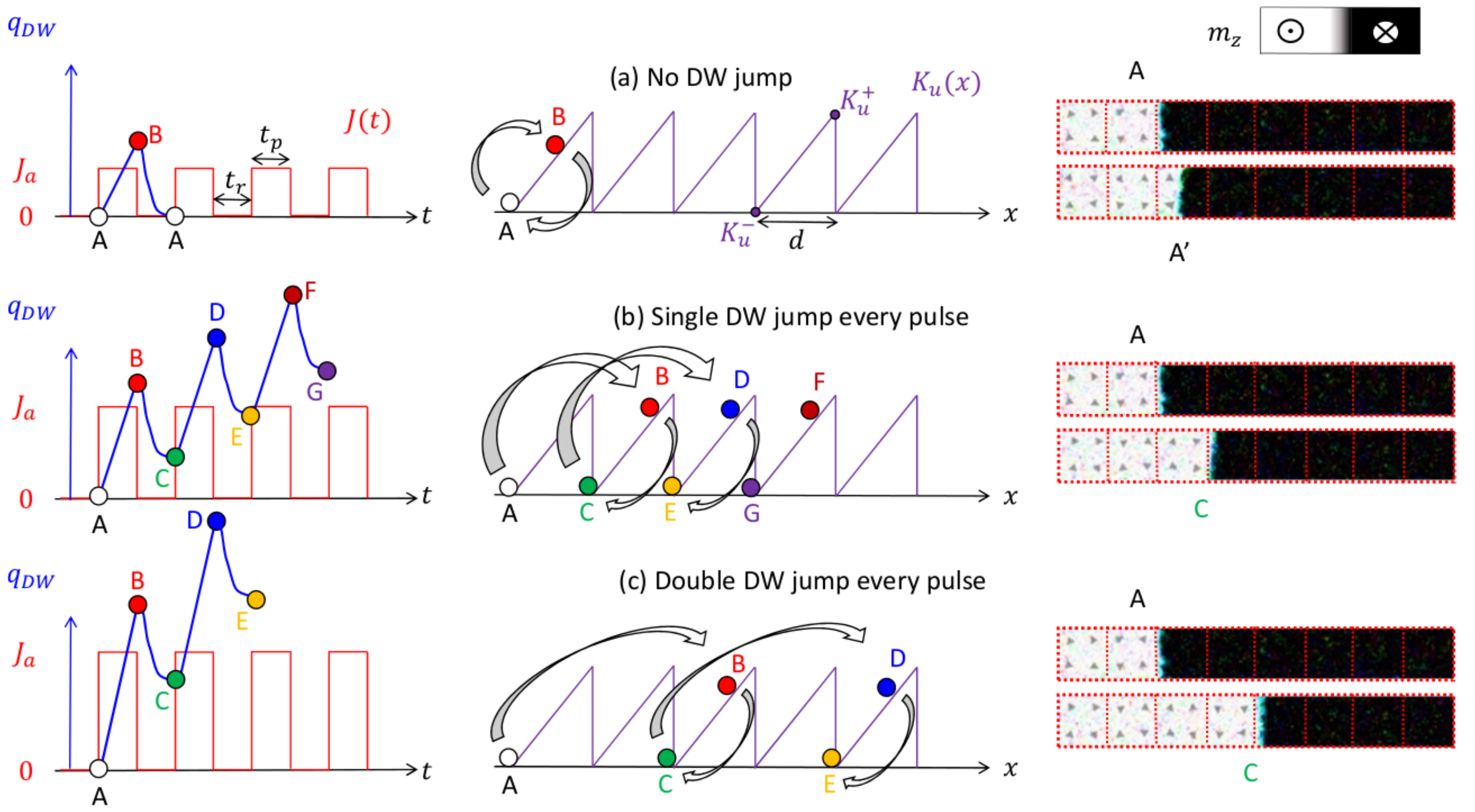}
\caption{DW motion under applied current in the FM ratchet considered. The sawtooth profile locates the DW in well defined areas as represented in the magnetization snapshots on the right. The final position of the DW depends on the current amplitude. If this amplitude is rather small, the DW does not cross any teeth and come back to its starting position A or an intermediate position A', as in the upper case. If the amplitude is excessively high, multiple teeth may be skipped at once, as in the lower case. The proper performance of the device is obtained when the DW skips one teeth at one current pulse.}\label{fig:outline}
\end{figure*}

A typical ratchet FM strip with high perpendicular magnetic anisotropy (PMA) sandwiched between a HM and an Ox is considered along the text. The ratchet anisotropy profile is obtained by tuning the out-of-plane anisotropy as done elsewhere. \cite{FrankenPinning} The central column of figure~\ref{fig:outline} depicts five periods of this anisotropy profile along the ratchet. Here the values of the highest and lowest anisotropy values, $K_u^+=1.27\frac{\mathrm{MJ}}{\mathrm{m}^3}$ and $K_u^-=1\frac{\mathrm{MJ}}{\mathrm{m}^3}$, respectively, are presented. A ratchet period $d=128$nm has been taken, as indicated also on the figure. The left column in figure~\ref{fig:outline} sketches the behaviour of a DW when short, intermediate, and large currents are applied. For low currents, the DW is not able to skip the teeth, labeled as point B in the upper subfigures, and then returns to its starting position A. Intermediate currents, however, lead the DW to skip one teeth at each pulse, i.e., a single DW jump, as it is depicted in the middle-right image. Finally, sufficiently large currents can make the DW skip two or more tooth at each pulse.\\

In order to check the operation range of this unidirectional ratchet device, material parameters commonly found in the literature \citep{Thia12} have been used for the present study: saturation magnetization $M_s$ of $1.1\frac{\mathrm{MA}}{\mathrm{m}}$, exchange constant $A$ of $16\frac{\mathrm{pJ}}{\mathrm{m}}$ and a Gilbert damping parameter $\alpha=0.5$. An interfacial DMI with a DMI parameter $D=1\frac{\mathrm{mJ}}{\mathrm{m}^2}$ has been considered. It must be noted that such an interfacial DMI is a constant, although the irradiation procedure is likely to affect not only the anisotropy, but also the DMI. However, to the best of our knowledge, there are no studies which link the change of the anisotropy constant with a change of the DMI value. Nevertheless, section~\ref{varDMI} has been included to show that a linear change of the DMI with a similar ratio to that considered for the PMA does not dramatically modify the results. The proposed DMI parameter is sufficiently high as to induce N\'{e}el DWs and so, to allow efficient current-driven DW movement by means of the SHE. Such an electric current is applied through the HM, being $\theta_{SH}=0.1$ the considered spin Hall angle. A FM strip with a cross section $L_y\times L_z$ of $128\mathrm{nm}\times 0.6\mathrm{nm}$ has been taken. The evolution in time of the system normalized magnetization $\vec{m}$ is described by the Landau-Lifshitz-Gilbert (LLG) equations augmented by the spin-orbit torques (SOT) and thermal fluctuations:
\begin{equation}\label{eq:thLLG}
\begin{split}
\frac{d\vec{m}}{dt}&={-\gamma_0}\vec{m}\times\left(\vec{H}_{eff}+\vec{H}_{th}\right)-\\
&{-\alpha}\vec{m}\times\frac{d\vec{m}}{dt}-\gamma_0\vec{m}\times\left(\vec{m}\times\vec{H}_{SH}\right)\mathrm{.}
\end{split}
\end{equation}
where $\gamma_0$ and $\alpha$ are the gyromagnetic ratio and the damping parameter, $\vec{H}_{eff}$ is the effective field, including exchange, anisotropy, and magnetostatic interactions, along with the DMI, and $\vec{H}_{SH}$ and $\vec{H}_{th}$ are, respectively, the effective field associated to the SHE and the thermal field, the latter included as a gaussian-distributed random field. \citep{BrownThermFluc, GarPalLang, EdMThermEff, DuineThCDDW} The study of the motion of a DW in such a system requires solving this equation by means of full micromagnetic simulations ($\mu$Mag). These $\mu$Mag simulations have been performed with the help of the Mumax$^{3}$ package. \citep{mumax}

\subsection{1DM}
The one dimensional model (1DM) was originally developed to describe the behaviour of one DW in an infinite wire, where magnetization changes only along the longitudinal coordinate. \citep{Thia02, Asym, Thia12} Within the framework of this model, two parameters account for the DW state, named $q$ and $\Phi$. The former determines the DW position, while the latter is, in our context, related to the relative orientation of the in-plane component of the magnetization with respect to the strip longitudinal direction. A further development of the 1DM \citep{FrankenPinning} contributed to adequately describe the behavior of one DW in an infinite strip with an anisotropy energy profile characterized by a linear variation in a region centred at $x=0$, and constant values of the anisotropy at the extremes of such a region. Here, a sawtooth profile, as that presented in figure~\ref{fig:outline}, is to be considered, i.e., a periodical repetition of a linear anisotropy profile with a period given by the previously defined distance $d$. Two mutually dependent equations can be derived for the DW dynamic behaviour in such a system:
\begin{subequations}\label{eq:th1DM}
\begin{align}
\begin{split}
\dot{q}&=\frac{\gamma_0\Delta}{1+\alpha^2}\left[\left(H_D{-H_k}\cos\Phi\right)\sin\Phi\right]+\\
&+\frac{\gamma_0\Delta}{1+\alpha^2}\left[\alpha\left( H_{SH}\cos\Phi+H_{th}+H_r r(q)\right)\right]\mathrm{,}
\end{split}
\\
\begin{split}
\dot{\Phi}&=\frac{\gamma_0}{1+\alpha^2}\left[\alpha\left(H_k\cos\Phi{-H_D}\right)\sin\Phi\right]+\\
&+\frac{\gamma_0}{1+\alpha^2}\left[H_{SH}\cos\Phi+H_{th}+H_r r(q)\right]\mathrm{.}
\end{split}
\end{align}
\end{subequations}
The terms $H_D$, $H_k$, $H_{SH}$ stand for the DMI, magnetostatic interaction, and SHE, respectively, and their definition can be found elsewhere. \citep{Thia12,Asym} Similarly, some bibliography can be posed where the thermal term $H_{th}$ is adequately discussed. \cite{BrownThermFluc, GarPalLang, EdMThermEff, DuineThCDDW} Finally, the anisotropy profile introduces the terms $H_r=\frac{K_u^+-K_u^-}{2\mu_0 M_s}$, and
\begin{equation}
\begin{split}
r(q)&=\frac{1}{\cosh^2\left[\left(1-\lbrace\frac{q}{d}\rbrace\right)\frac{d}{\Delta} \right]}+\frac{1}{\cosh^2\left(\lbrace\frac{q}{d}\rbrace\frac{d}{\Delta} \right)}-\\
&-\frac{ \frac{\Delta}{d} \sinh \left(\frac{q}{\Delta}\right)} {\cosh\left(\lbrace\frac{q}{d}\rbrace\frac{d}{\Delta}\right)\cosh\left[\frac{\left(1-\lbrace\frac{q}{d}\rbrace\right)d}{\Delta} \right]}\mathrm{,}
\end{split}
\end{equation} 
where the braces stand for the fractional part function. It must be pointed out here that well-defined DW equilibrium positions can be derived from (\ref{eq:th1DM}), which are found close to the edges of the sawtooth profile, where the anisotropy minima are located. More details are given in Appendix~\ref{AppA}.

\section{Results}\label{Results}
\subsection{Comparison between the 1DM and full $\mu$Mag simulations at $300K$ for a $50\%$ duty cycle}\label{50C}

\begin{figure}[t!]
\includegraphics[width=\columnwidth]{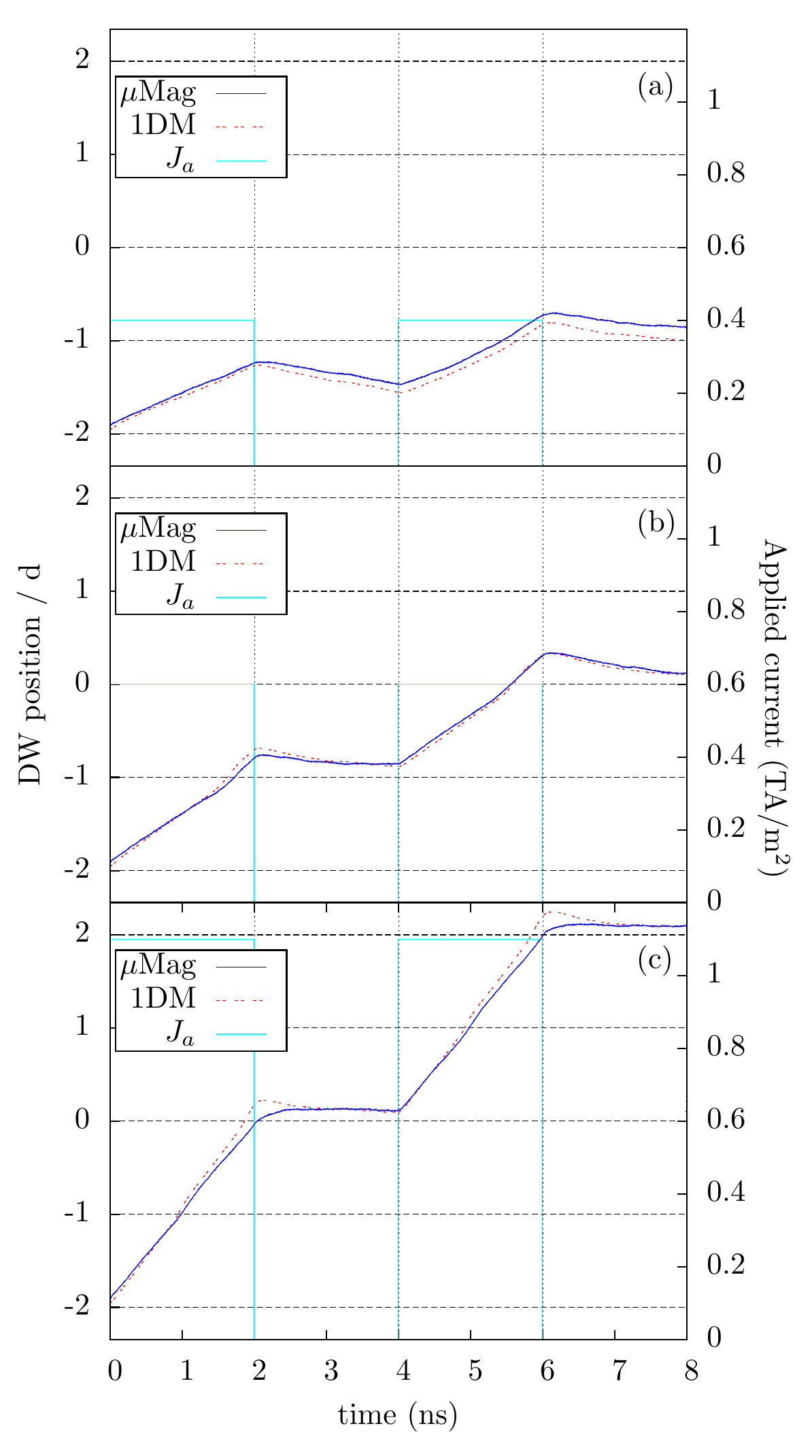}
\caption{DW position as a function of time for a 2-ns pulse length with a 2-ns relaxation time. Three different current amplitudes are considered: (a) $J_a=0.4\frac{\mathrm{TA}}{\mathrm{m}^2}$, (b) $J_a=0.6\frac{\mathrm{TA}}{\mathrm{m}^2}$, and (c) $J_a=1.1\frac{\mathrm{TA}}{\mathrm{m}^2}$. The graphs also compare the results drawn by both $\mu$Mag simulations (solid lines) and the 1DM (dashed lines). The vertical black dashed lines indicate the time at which HM current is switched on and off, while the horizontal black dashed lines mark the subsequent positions of each teeth (${-256}$, ${-128}$, $0$ and $128$, $256$nm, for $d=128$nm).}\label{fig:TemEv}
\end{figure}

The working principle for the device presented in section~\ref{GeoMod} is the following. Each region of length $d$ can pin one DW, i.e., the region size defines the bit size. When a pulse of current is applied to the system, every DW moves forward during the pulse time $t_e$. When the current stops, the anisotropy interaction drives the DW during a relaxation time $t_r$ towards the closest anisotropy minimum, where the DW is more stable. If the amplitude of the pulse is rather low, the DW does not skip any tooth of the anisotropy landscape, and no bit shifting occurs, as the image on top of figure~\ref{fig:outline} depicts. A certain correspondence between this figure and the graph labelled as figure~\ref{fig:TemEv}(a) can be checked. Actually, the graph in this figure reflects the case when the DW is firstly driven by the current for a time $t_e$ a distance shorter than $d$, up to a point B. The DW does not skip the anisotropy tooth and reverses during the time $t_r$. If $t_r$ is sufficiently long, the DW recovers its starting position A at equilibrium. However, the time $t_r$ is in this case rather short, and the DW acquires an intermediate position A' when the current is switched on again. An eventual bit shifting may occur during the application of the second current pulse, since the DW may skip the anisotropy tooth from such an intermediate position. This is, in general, an undesirable behavior. Contrarily, if the amplitude of the pulse is adequately high as to displace the DW a distance larger than $d$ in a time $t_e$, the DW skips the anisotropy tooth and does not return to its starting position, but to the subsequent equilibrium position, a distance $d$ away from its starting position, if the relaxation time is sufficiently long. This can be seen in figure~\ref{fig:TemEv}(b), which corresponds to the central case of figure~\ref{fig:outline}, being this situation the desired behavior, since the information is shifted only one bit at a time, i.e., a single DW jump occurs. Finally, if the current amplitude is rather high, the DW may reach an intermediate position prior to the application of the next pulse. This can cause the DW to eventually advance two bits at a time, instead of only one, when subsequent pulses are applied. What is more, for sufficiently high currents, the DW runs a distance of two bits at once instead of only one, as figure~\ref{fig:TemEv}(c) depicts, which corresponds to the bottom case in fig~\ref{fig:outline}.\\

\begin{figure}[t]
\includegraphics[width=\columnwidth]{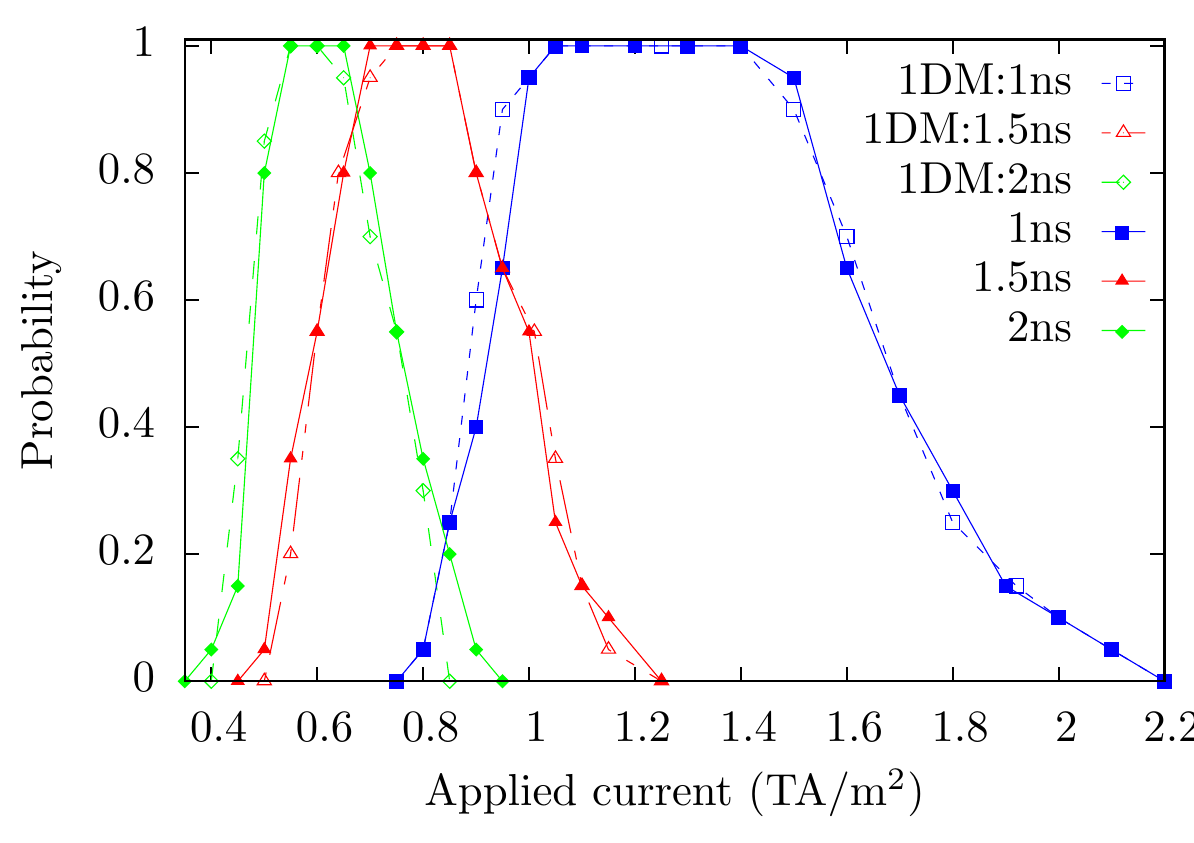}
\caption{Probability of single jumps of one DW after the application of one pulse with a duty cycle of 50\% as a function of the current amplitude for pulses with $t_e=1\mathrm{ns}$, $1.5\mathrm{ns}$, and $2\mathrm{ns}$, computed from full $\mu$Mag simulations (filled symbols) and the 1DM (open symbols).}\label{fig:LengthPulse}
\end{figure}

The proper range of operation of the device is the range of values of the current amplitude $J_a$ for a given pair of times $t_e$ and $t_r$ that promote single DW jumps after the application of one current pulse. As an example, Figure~\ref{fig:LengthPulse} depicts the probability of one single DW jump after the application of one pulse as a function of the current, obtained from both $\mu$Mag simulations and the 1DM, when $t_e=t_r$, that is, a duty cycle of 50\% for the injected current. Three different pulse times, $t_e=1\mathrm{ns}$, $1.5\mathrm{ns}$ and $t_e=2\mathrm{ns}$, are considered. The probability has been statistically obtained by evaluating twenty different realizations for each current amplitude and pulse length. It must be noticed that the proper range of operation for a duty cycle of $50\%$ increases with decreasing pulse lengths. Alternatively, next subsection shows that a more efficient way to increase the range of operation is to increase only the relaxation time, then varying the duty cycle. In any case, it can be checked from the plots in this graph that the results provided by full $\mu$Mag simulations and the 1DM are in a rather good agreement.

\subsection{Comparison between one dimensional model and full $\mu$Mag simulations at $300K$ for $33\%$ duty cycle}\label{33C}

With the aim of extending the 100\% probability range of proper working of this device, the following strategy can be proposed. Since the DW requires a minimal relaxation time to reach an equilibrium position, this time can be lengthened, then reducing the current duty cycle. As an example, the following results refer to a ratchet memory initially driven by an injected current with a time pulse $t_e=1\mathrm{ns}$, and further relaxing for $t_r=2\mathrm{ns}$ prior to the application of the subsequent pulse, that is, the duty cycle is reduced to 33\%.

\begin{figure}[t]
\includegraphics[width=\columnwidth]{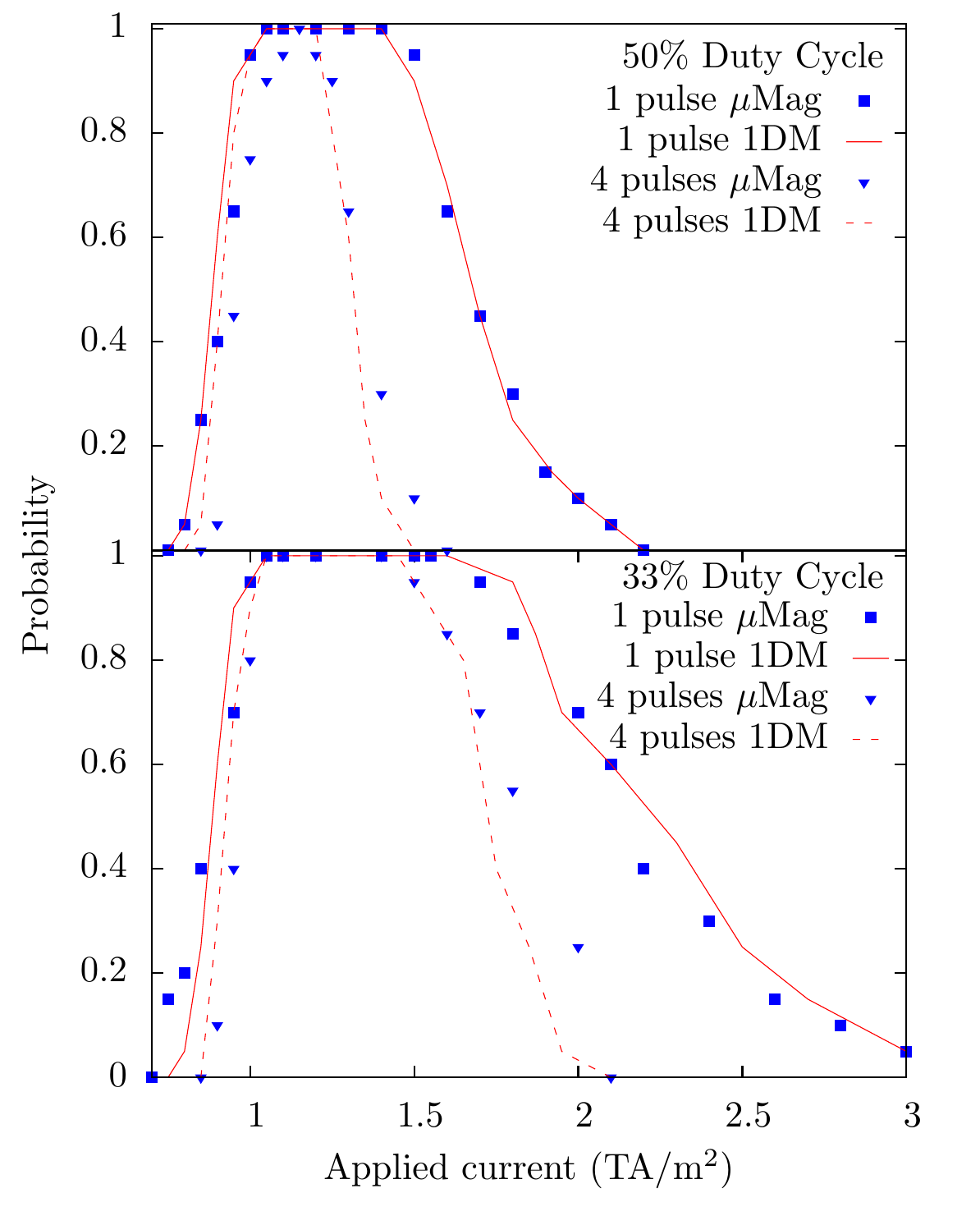}
\caption{Probability of one single jump of a DW promoted by one pulse current with $50\%$ (top graph) and $33\%$ (bottom graph) duty cycles (solid line for the results provided by the 1DM and squares for those obtained from $\mu$Mag simulations), and probability of four single jumps after four current pulses (dashed line for the results provided by the 1DM and triangles for those obtained from $\mu$Mag simulations). The pulse time is $t_e=1\mathrm{ns}$ in both cases, while the relaxation times are $t_r=1\mathrm{ns}$ for the $50\%$ duty cycle and $t_r=2\mathrm{ns}$ for the $33\%$ duty cycle.}\label{fig:CompCycle}
\end{figure}

Figure~\ref{fig:CompCycle} compares the probability of single jumps of a DW after one current pulse (one single jump), and after four pulses (four consecutive single DW jumps) calculated by means of $\mu$Mag simulations and the 1DM. Two duty cycles, $50\%$ and $33\%$, are considered.

As in the preceding case, twenty different realizations have been carried out in order to compute statistics. The results for current pulses with a $50\%$ duty cycle (see top graph) reveal that the range of proper working of the device notably reduces as the number of pulses is increased from one to four (the latter may be regarded as the fourth power of the after-one-pulse results). This fact makes the system much less reliable when long bit shifting is needed, or when a long array of bits must be displaced. However, the bottom graph in the figure reveals that, even though the range of proper operation also reduces as the number of pulses is increased, this range is still rather wide in the case of a $33\%$ duty cycle. Thus, it is possible to improve the reliability of the current-driven ratchet memory by lowering the current duty cycle.\\

\begin{figure}[t]
\includegraphics[scale=0.44]{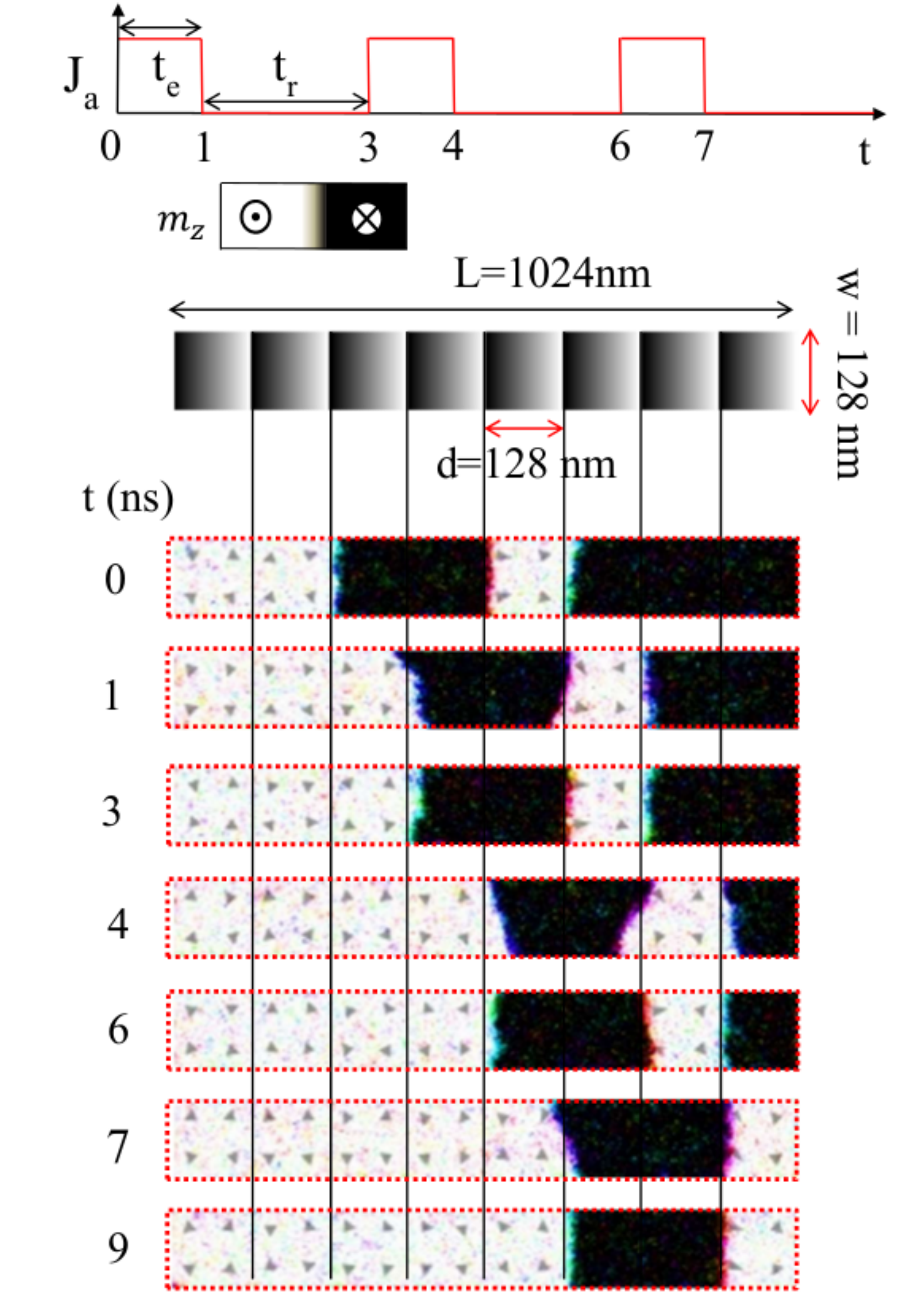}
\caption{The operation of a current-driven ratchet memory containing eight bits is studied by means of $\mu$Mag simulations. Three $33\%$ duty cycle current pulses of width $t_e=1\mathrm{ns}$ and relaxation times $t_r=2\mathrm{ns}$ are applied to the device. The current amplitude is fixed to $J_a=1.2\frac{\mathrm{TA}}{\mathrm{m}^2}$. The intermediate figure displays the anisotropy landscape, which defines the bit size, as a color map. The storage of every bit then requires a surface of $d\times w=128\times 128\mathrm{nm}^2$, to complete a total length of the ratchet of $L=1024\mathrm{nm}$. Snapshots at the end of every subsequent interval of time either $t_e$ or $t_r$ are sketched to illustrate the dynamical behaviour.}\label{fig:Dynamic}
\end{figure}

Figure~\ref{fig:Dynamic} depicts the magnetization of an array of eight bits in a ratchet memory with a bit length given by the period $d=128\mathrm{nm}$. Three $33\%$ duty cycle current pulses of width $t_e=1\mathrm{ns}$ and relaxation times of $t_r=2\mathrm{ns}$ are applied to the device. The figure demonstrates the correct operation of the system, which drastically improves the performances of the field-driven ratchet memory. \citep{RatchMem} In particular the bit size is reduced to an approximate factor of two. In fact, the field-driven system requires the length of two periods of the anisotropy landscape to store one bit, while one period is needed to trap every DW in the case of the current-driven device, then reducing the bit size to the length of only one period. Furthermore, Walker field limits the highest DW speed in the case of the field-driven system. The required time to shift one bit is in that case as long as $16\mathrm{ns}$. However, $3\mathrm{ns}$ are sufficient to carry out one correct bit shifting in the case of the current-driven element here proposed.\\

\subsection{Effect of a variable DMI}\label{varDMI}
\begin{figure}[t]
\includegraphics[width=\columnwidth]{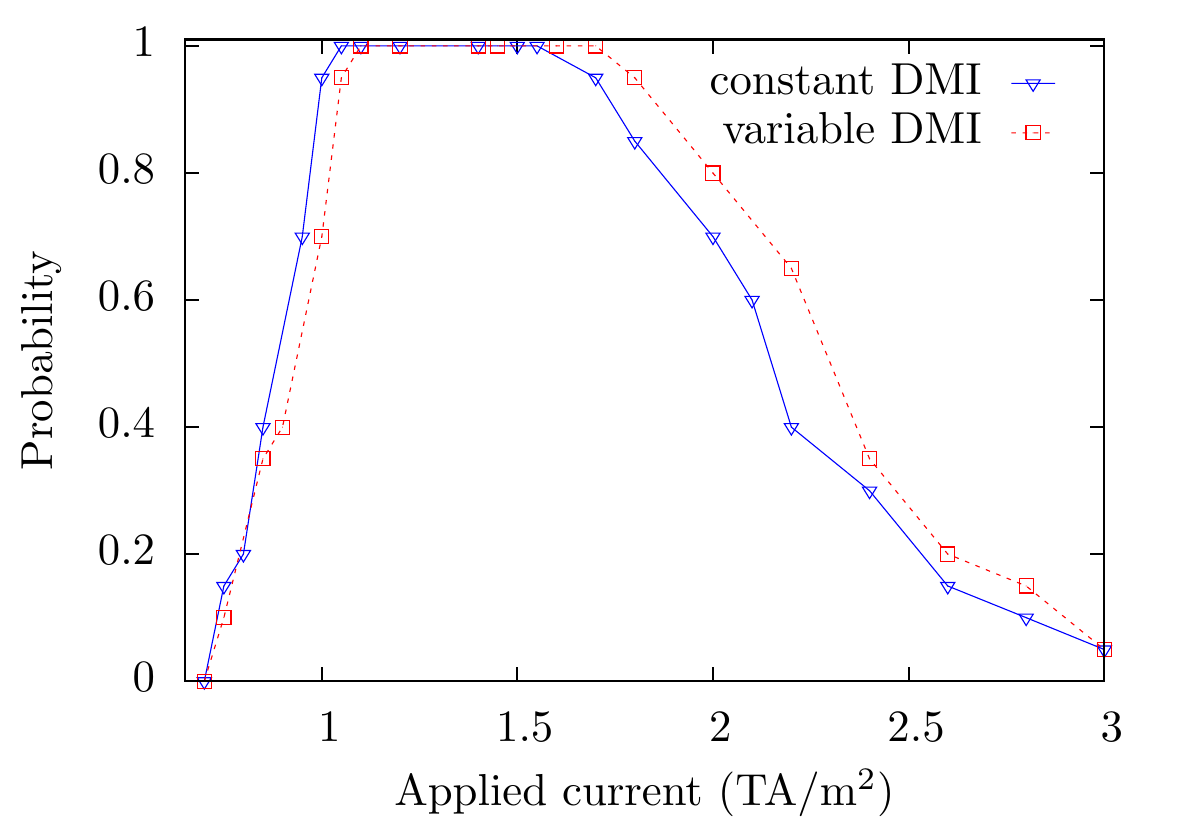}
\caption{Probability of single jumps of one DW for the model with constant DMI parameter (solid line) and with a linear variation of the DMI parameter (dashed line).}\label{fig:CompDMI}
\end{figure}

Finally, it must be reminded that a constant DMI parameter has been considered along this work. Although this assumption might not seem sufficiently realistic, to the best of our knowledge there are no studies on how the DMI changes when the HM-FM interface is irradiated with heavy ions so as to tailor the PMA. As a first approach to the effect of this irradiation on the DMI, a linear variation of the DMI parameter, analogous to that of the PMA, can be proposed. In this way, $\mu$Mag simulations have been carried out to compare the results obtained both on the constant DMI basis and from the proposed linear variation of the DMI. In the latter case, the DMI is considered to change between $D_-=0.8\frac{\mathrm{mJ}}{\mathrm{m}^2}$ and $D_+=1\frac{\mathrm{mJ}}{\mathrm{m}^2}$. 33\% duty cycle current pulses are applied with an amplitude $J_a=1.2\frac{\mathrm{TA}}{\mathrm{m}^2}$, and the probability of single jumps is analysed in both cases, then drawing the results plotted in figure~\ref{fig:CompDMI}. No drastic changes are found in the case of variable DMI with respect to the case of constant DMI, and the predictions about the proper range of operation are not highly affected.\\

\section{Conclusions}\label{concl}

The work presents a study of the current-driven ratchet memory as a DW-based magnetic memory. This study has been carried out by means of full $\mu$Mag simulations along with the results provided by a 1DM, showing both approaches a rather good agreement. It was necessary in the latter model to introduce a new term in the time derivatives of the DW position $q$ and the magnetization angle $\Phi$ taking into account the space dependent anisotropy profile which makes it possible this one-way working principle. The work shows how bit density and shifting speeds are notably increased if the memory device works on this current-driven basis, as compared with the field-driven scheme. A fine tuning of the pulse and relaxation times may even improve its performances. In this way, the best results are obtained when the combined effect of the times of both the applied current pulse and the relaxation interval leads every DW from an equilibrium state to another one at the contiguous tooth of the anisotropy landscape, the former time by promoting only one single jump and the latter by allowing the DW to recover this new equilibrium position.\\

\renewcommand{\theequation}{\thesection.\arabic{equation}}
\setcounter{equation}{0}

\appendix
\section{Derivation of the one dimensional model}\label{AppA}

In order to derive the 1DM the usual assumption that magnetization only changes appreciably in one direction is considered. The well known Walker ansatz is taken in order to describe the magnetization along the FM strip where a DW is formed. The use of variational principles \citep{Thia02, WalkerBrd} leads in the case of a strip with constant anisotropy to a couple of well established equations for the DW position and the magnetization orientation within the DW. In our case, and due to the additive character of the energy accounting for the considered interactions, the only term in these equations that must be explicitly rewritten is the one resulting from the integration of the anisotropy landscape along the longitudinal direction. Accordingly, the expression of the magnetocristalline anisotropy density energy $u_{anis}=K_u \sin^{2} \theta$ is to be considered along this section, where $\theta$ is the angle between the magnetization and the out-of-plane direction ($z$). As to draw some preliminary results, the case of a single anisotropy slope can be recalled. \citep{FrankenPinning} In such a case $K_u$ is no longer a constant, and is defined as:
\begin{equation}
K_u\left(x\right) = \left\lbrace \begin{array}{ccc}
K_u^- &\mathrm{if}  &x<-\frac{d}{2}\mathrm{,}\\
\frac{K_u^+ + K_u^-}{2}+\frac{K_u^+ - K_u^-}{d}·x  &\mathrm{if}  &-\frac{d}{2}<x< \frac{d}{2}\mathrm{,}\\
K_u^+ &\mathrm{if} &x>\frac{d}{2}\mathrm{,}
\end{array} \right.
\end{equation}
where $K_u^+$ is the uniaxial anisotropy constant at the right side of the strip while $K_u^-$ is the uniaxial anisotropy constant at the left side, $d$ being the distance of variation of the anisotropy. If the Walker ansatz is considered to define the magnetization along the FM strip, i.e., $\theta=2\arctan\left[\exp \left( \frac{x-q}{\Delta} \right) \right]$, $q$ being the DW position and $\Delta$ the DW width, taken as a constant value, then the anisotropy free energy density per unit surface $\sigma_{anis}$ is given by:
\begin{equation}\label{eq:sum}
\begin{split}
\sigma_{anis}&=\int_{-\infty}^{\infty} K \sin^{2}\theta dx=\\
&=\int_{-\infty}^{-\frac{d}{2}} K_u^- \sin^{2}\theta dx +\\
&+ \int_{-\frac{d}{2}}^{\frac{d}{2}} \left( \frac{K_u^+ + K_u^-}{2}+\frac{K_u^+ - K_u^-}{d}·x \sin^{2}\theta \right)dx +\\
&+ \int_{\frac{d}{2}}^{\infty} K_u^+ \sin^{2}\theta dx\mathrm{,}
\end{split}
\end{equation}
where the first and third integral can be obtained as:
\begin{equation}
\begin{split}
I_1&=\Delta K_u^-\left[1-\tanh\left(\frac{\frac{d}{2}+q}{\Delta}\right) \right]\mathrm{,}\\
I_3&=\Delta K_u^+\left[1-\tanh\left(\frac{\frac{d}{2}-q}{\Delta}\right) \right]\mathrm{,}
\end{split}
\end{equation}
and the second integral is:
\begin{equation}
\begin{split}
I_2&=\Delta \frac{K_u^+ +K_u^-}{2}\left[\tanh\left(\frac{\frac{d}{2}-q}{\Delta}\right)+\tanh\left(\frac{\frac{d}{2}+q}{\Delta}\right) \right]+\\
&+\Delta\frac{K_u^+ - K_u^-}{2}\left[\tanh\left(\frac{\frac{d}{2}-q}{\Delta}\right)-\tanh\left(\frac{\frac{d}{2}+q}{\Delta}\right)\right]+\\
&+\left(K_u^+ - K_u^-\right)\frac{\Delta^{2}}{d}\ln\left[\frac{\cosh\left(\frac{\frac{d}{2}+q}{\Delta}\right)}{\cosh\left(\frac{\frac{d}{2}-q}{\Delta}\right)}\right]\mathrm{.}
\end{split}
\end{equation}
By adding up this results, (\ref{eq:sum}) can be worked out as:
\begin{equation}\label{eq:saniss}
\begin{split}
\sigma_{anis}&=\Delta\left(K_u^+ + K_u^-\right) + \\
&+\frac{\Delta^{2}}{d}\left(K_u^+ - K_u^- \right)\ln\left[\frac{\cosh\left(\frac{\frac{d}{2}+q}{\Delta}\right)}{\cosh\left(\frac{\frac{d}{2}-q}{\Delta}\right)}\right]\mathrm{.}
\end{split}
\end{equation}
which leads to the following interaction term as due to the local variation of the anisotropy:
\begin{equation}
\frac{\partial\sigma_{anis}}{\partial q}=\frac{\Delta}{d}\left(K_u^+ - K_u^- \right)\frac{2\sinh\left(\frac{d}{\Delta}\right)}{\cosh\left(\frac{d}{\Delta} \right)+\cosh\left(\frac{2q}{\Delta}\right)}
\end{equation}
The effect of this term vanishes for the constant anisotropy areas $\left(\left|q\right|>\frac{d}{2}\right)$), and drives the wall towards lower anisotropy areas when a local variation of this interaction appears, as expected. Accordingly, for an anisotropy variation following a sawtooth profile, i.e., an anisotropy slope periodically repeated as shown in figure~\ref{fig:outline}, the local anisotropy minima define equilibrium positions for the DWs. This can be analytically demonstrated by taken an anisotropy parameter $K_u\left(x\right)$ defined by the formula:
\begin{equation}
K_u\left(x\right)= K_u^-+\frac{K_u^+ - K_u^-}{d}·\left[x-\lfloor\frac{x}{d}\rfloor d\right]\mathrm{,}
\end{equation}
where the brackets $\lfloor\:\rfloor$ stand for the floor function. For the sake of simplicity, $K_u\left(x\right)$ has been considered to be discontinuous between the points of maximal and minimal anisotropy. $\sigma_{anis}$ can be worked out as:
\begin{equation}
\sigma_{anis}=\int_{0}^{L} K_u\left(x\right)\sin^{2}\theta dx=\sum_{n=0}^{N-1} I_n\mathrm{,}\label{eq:A8}
\end{equation}
$L$ being the total length of the ratchet, then containing $N$ slopes, that is, $L=Nd$. $I_n$ is the result of the integral of the anisotropy along the $n$-th slope, $n$ enumerating the subsequent slopes, and ranging from 0 to $N-1$. $I_n$ then reads:

\begin{equation}
\begin{split}
I_n&=\Delta K_u^-\tanh\left[\frac{\left(\lbrace \frac{q}{d}\rbrace+n\right)d}{\Delta}\right]-\\
&-\Delta K_u^+\tanh\left[\frac{\left(\lbrace \frac{q}{d}\rbrace+n{-1}\right)d}{\Delta}\right]+\\
&+\frac{\Delta^2}{d}\left(K_u^+-K_u^-\right)\ln\left[\frac{\cosh\left[\frac{\left(\lbrace \frac{q}{d}\rbrace+n\right)d}{\Delta}\right]}{\cosh\left[\frac{\left(\lbrace \frac{q}{d}\rbrace+n{-1}\right)d}{\Delta}\right]}\right]\mathrm{.}
\end{split}
\end{equation}
It must be reminded here that the braces stand for the fractional part function. Under the assumption that the DW width $\Delta$ is negligibly small if compared with the length $d$, the sum in (\ref{eq:A8}) adds up to:\\
\linebreak
\pagebreak
\begin{equation}\label{eq:sanisr}
\begin{split}
\sigma_{anis}&=\Delta\left( K_u^+ + K_u^-\right)+\\
 &+\Delta \left( K_u^+ + K_u^-\right)\tanh\left[\left(1{-\lbrace\frac{q}{d}\rbrace}\right)\frac{d}{\Delta} \right] -\\
 &-\Delta \left( K_u^+ + K_u^-\right)\tanh\left(\lbrace\frac{q}{d}\rbrace\frac{d}{\Delta} \right)+\\
 &+\frac{\Delta^2}{d}\left( K_u^+ + K_u^-\right)\ln\left[\frac{\cosh\left(\lbrace\frac{q}{d}\rbrace\frac{d}{\Delta} \right)}{\cosh\left[\frac{\left(1-\lbrace\frac{q}{d}\rbrace\right)d}{\Delta} \right]}\right]\mathrm{.}
\end{split}
\end{equation}

This result, together with the other contributions to the energy density per unit surface, as long as the thermal field,\cite{BrownThermFluc, GarPalLang, EdMThermEff, DuineThCDDW} allows us to finally write the 1DM equations (\ref{eq:th1DM}).

\bibliography{\jobname}

\end{document}